\documentclass[sigconf]{acmart}
\usepackage{enumerate}
\usepackage{multirow}
\usepackage{threeparttable}
\usepackage{xcolor}
\usepackage{listings}
\usepackage{graphicx}
\usepackage{tikz,lipsum,lmodern}
\usepackage[most]{tcolorbox}
\usepackage{caption}

\graphicspath{ {./images/} }
\definecolor{mGreen}{rgb}{0,0.6,0}
\definecolor{mGray}{rgb}{0.5,0.5,0.5}
\definecolor{mPurple}{rgb}{0.58,0,0.82}
\definecolor{backgroundColour}{rgb}{0.95,0.95,0.92}

\lstdefinestyle{CStyle}{
    backgroundcolor=\color{backgroundColour},   
    commentstyle=\color{mGreen},
    keywordstyle=\color{magenta},
    numberstyle=\tiny\color{mGray},
    stringstyle=\color{mPurple},
    basicstyle=\footnotesize,
    breakatwhitespace=false,         
    breaklines=true,                 
    captionpos=b,                    
    keepspaces=true,                 
    numbers=left,                    
    numbersep=5pt,                  
    showspaces=false,                
    showstringspaces=false,
    showtabs=false,                  
    tabsize=2,
    language=C
}

\acmConference[SIGCSE Virtual 2024]{1st ACM Virtual Global Computing Education Conference}{December 2024}{Virtual}

\AtBeginDocument{%
  \providecommand\BibTeX{{%
    \normalfont B\kern-0.5em{\scshape i\kern-0.25em b}\kern-0.8em\TeX}}}

\usepackage{hyperref} 
\begin{document}

\title{Integrating AI Tutors in a Programming Course}


\author{Iris Ma}
\email{huaiyaom@uci.edu}
\affiliation{%
 \institution{University of California, Irvine}
 \city{Irvine}
 \state{California}
 \country{USA}
}

\author{Alberto Krone Martins}
\email{algol@uci.edu}
\affiliation{%
 \institution{University of California, Irvine}
 \city{Irvine}
 \state{California}
 \country{USA}
}

\author{Cristina Videira Lopes}
\email{lopes@uci.edu}
\affiliation{%
 \institution{University of California, Irvine}
 \city{Irvine}
 \state{California}
 \country{USA}
}

\begin{abstract}
    RAGMan is an LLM-powered tutoring system that can support a variety of course-specific and homework-specific AI tutors. RAGMan leverages Retrieval Augmented Generation (RAG), as well as strict instructions, to ensure the alignment of the AI tutors' responses. By using RAGMan's AI tutors, students receive assistance with their specific homework assignments without directly obtaining solutions, while also having the ability to ask general programming-related questions.

RAGMan was deployed as an optional resource in an introductory programming course with an enrollment of 455 students. It was configured as a set of five homework-specific AI tutors. This paper describes the interactions the students had with the AI tutors, the students' feedback, and a comparative grade analysis. Overall, about half of the students engaged with the AI tutors, and the vast majority of the interactions were legitimate homework questions. When students posed questions within the intended scope, the AI tutors delivered accurate responses 98\% of the time. Within the students used AI tutors, 78\% reported that the tutors helped their learning. Beyond AI tutors' ability to provide valuable suggestions, students reported appreciating them for fostering a safe learning environment free from judgment.

\end{abstract}

\keywords{large language models, education, software engineering, LLMs}


\maketitle

\section{Introduction}
\label{sec:intro}

The number of students enrolling in Computer Science has increased significantly in the past decades. This surge is driven by the growing reliance on software in modern society and the increasing appeal of software careers. We teach in a large public University in the United States, and our undergraduate enrollment in Computer Science has reached peak levels, with close to 1,000 freshmen every year over the past 3 years. Providing individualized support to a large number of students in introductory programming classes, especially regarding mastery of challenging material, has been challenging.

The emergence of conversational chatbots built on top of Large Language Models (LLMs) presents a potential route for providing students with nteractive learning experiences and on-demand, personalized assistance with their homework assignments. We developed RAGMan, a Retrieval-Augmented Generation (RAG) framework for building AI tutors, and deployed several of these tutors in an introductory programming course, ICS32. The AI tutors were configured so that they would not give out the solutions to the homework, and would nudge the students towards possible next steps relative to their questions. We collected anonymized usage data, and asked the students to fill out a survey at the end. Additionally, we had access to the anonymized final grades.

The data from this deployment points to a positive outcome regarding the benefit, and perceived benefit, of the AI tutors: there were statistically significant fewer students who failed the course compared to the previous year's cohort, and the vast majority of students who chose to use the AI tutors reported that they felt they helped their learning. But there are also some warning signs, specifically: there was a very small, but marginally significant decrease in the percentage of A-grades. As this is just one deployment, and not a controlled or systematic experiment, we cannot assert causation, or even correlation, between the use of AI tutors and the changes in grades. Like many recent uses of AI tutors, ours is exploratory in nature; its value is to help pave the way for future research.

We believe that we are entering a new world in Education. Only 21\% of students in this cohort reported not using AI tools at all for homework help, and 56\% of them reported using several AI tools. On the one hand, the use of unconstrained AI tools, such as ChatGPT and Copilot, threatens to disrupt the human learning process, as they can simply give out answers to homework questions, bypassing the student's productive struggles. On the other hand, it is now possible to develop educational tools that help the students in their learning process. Knowing where to draw the lines is an emerging art.



\section{Background}
\label{sec:background}

\subsection{LLMs in Education}


Large Language Models (LLMs) refer to extensive deep learning models constructed on the foundation of the transformer architecture~\cite{vaswani2017attention} and pre-trained with a substantial volume of data. LLMs have proven successful in solving a variety of Natural Language Processing (NLP) tasks~\cite{madaan-etal-2022-language, STEISS2024writing, macneil2023code-explain, leinonen2023comparing-code-explain} and passing knowledge examinations crossing various fields~\cite{katz2024gpt-bar, wang2023medical-exam, biswas2023role-cs}. Given the versatility of LLMs in solving problems across different domains, there has been a growing interest in using LLMs in education based on their potential to improve learning experiences, enable personalized education, and help with creating and grading assignments.~\cite{kasneci2023chatgpt-good}.

\subsubsection{Impacts of Using LLMs in Educational Settings} 
Several studies have explored the opportunities and challenges associated with integrating AI tools into education ~\cite{warschauer2023affordances, Kasneci2023challenges,becker2023cs-education, 10.1145/3626253.3635604}. Jacob et al.~\cite{jacob2023gpt-writing}  investigated the experience of a non-native English speaker using ChatGPT to improve their English writing skills. The authors found that such tools can positively impact second-language learners by facilitating writing practice, rather than hindering their development.

While AI tools powered by LLMs offer potential benefits for student learning, concerns regarding their inclusion, usability, technical limitations, ethical implications, and user well-being have also been raised~\cite{durall2020bot-concerns}. A study~\cite{tlili2023gpt-devil} revealed that ChatGPT could exhibit harmful behaviors like dishonesty, manipulation, and misinformation dissemination. 
Thus, the development of responsible and accountable AI tools for real-world applications is critical~\cite{kuhail2023bot-design, arrieta2020xai}.

\subsubsection{LLMs in Computer Science Education}
Research on the application of LLMs in computer science education is still in its early stages. Studies have primarily focused on exploring their capabilities~\cite{kazemitabaar2024codeaid}, particularly in programming task solving~\cite{denny2023conversing, Finnie-Ansley2023codex-task} and educational content generation ~\cite{leinonen2023edu-content}. 

Three recent studies have implemented LLM-powered AI tools for programming classes; all of these works are similar to ours, although with significant differences. 

Liffiton et al.~\cite{liffiton2023codehelp} developed CodeHelp for entry-level programming classes. CodeHelp presents the students a fixed number of forms with which they can get help about specific blocks of code -- explanations, error messages, etc. It was designed with a number of guardrails that ensure that the complete solution to programming problems is not presented to the students, and it only responds to content related to programming. Also, code blocks are removed from the responses. In order to implement these guardrails, CodeHelp makes several requests to the LLM, with several prompts.The results indicate that students found CodeHelp to be a valuable complement to direct support from instructors and teaching assistants.

Kazemitabaar et al.~\cite{kazemitabaar2024codeaid} developed CodeAid, a system that leverages few-shot learning to assist students with programming assignments and enhance their conceptual understanding. CodeAid is very similar to CodeHelp, in that it presents the students with a fixed set of help actions. Their prompts include examples (few-shot learning) for improved responses. Their results indicate that the tool provided correct and helpful answers in \~90\% of the time.

Liu et al.~\cite{cs50ai2024} introduced a suite of AI-powered educational tools within a web application named CS50.ai, designed to enhance learning experiences in an introductory programming course. These tools provide functionalities including code explanation, style checking, and a Q\&A bot integrated with Ed Discussion~\footnote{\url{https://edstem.org/}}). The Q\&A bot employs the RAG technique, achieving an accuracy of 88\% in responding to curriculum-related queries and 77\% in administrative queries. 

Our AI tutors differ from both CodeHelp and CodeAid in a couple of important aspects. First, our AI tutors are conversational agents, not single question help tools. We wanted to provide free-text conversational AI tutors whose user interface is similar to a human Teaching Assistant available over UIs such as Slack. This is because, often, students do not know enough about their questions to be able formulate them in the right way, or even classify them properly; having a conversation with an AI tutor has the potential to help students discover what their question really is, and to engage with the student's questions in a much deeper manner. Second, internally, our RAGMan framework uses the RAG technique, which allows us to scope the responses to the exact topics of the homework assignments.

As for CS50.ai, the difference is that we have implemented the RAG technique in the conversations between students and AI tutors instead of using it to answer administrative questions on Ed. Our study also analyzed these interactions by evaluating the accuracy of responses generated by the AI tutors using the RAG technique. This analysis not only sheds light on the effectiveness of the RAG application in conversational educational settings but also provides insights into potential areas for further enhancement of AI-driven educational tools.



\subsection{Retrieval-Augmented Generation (RAG)}

When users want to explore detailed knowledge into current or specific topics with an LLM, they may encounter responses that are hallucinatory and misleading~\cite{ji2023-hallucination,alkaissi2023gpt-hallucination, li2023halueval-hallucination,lin2024assessment-rag}. According to Li et al.~\cite{li2023halueval-hallucination}, about 19.5\% responses from ChatGPT are hallucinatory. To mitigate this, Lewis et al. have developed RAG, which connects generative AI models to external knowledge database~\cite{lewis2020RAG}. RAG equips models with sources they can reference to make their responses verifiable and more dependable.

\section{RAGMan and AI Tutors}
\label{sec:teach-arch}


RAGMan is our RAG framework for developing AI tutors, and it is fairly generic -- it can be used for education as well as other purposes. Each AI tutor built on top of  RAGMan has its own knowledge base, and behaves in ways that are tuned for specific goals that are not necessarily all the same for all AI tutors. Here we explain the generic framework, and how we obtain specific AI tutors from it.

\begin{figure}
    \centering
    \includegraphics[width=.4\textwidth]{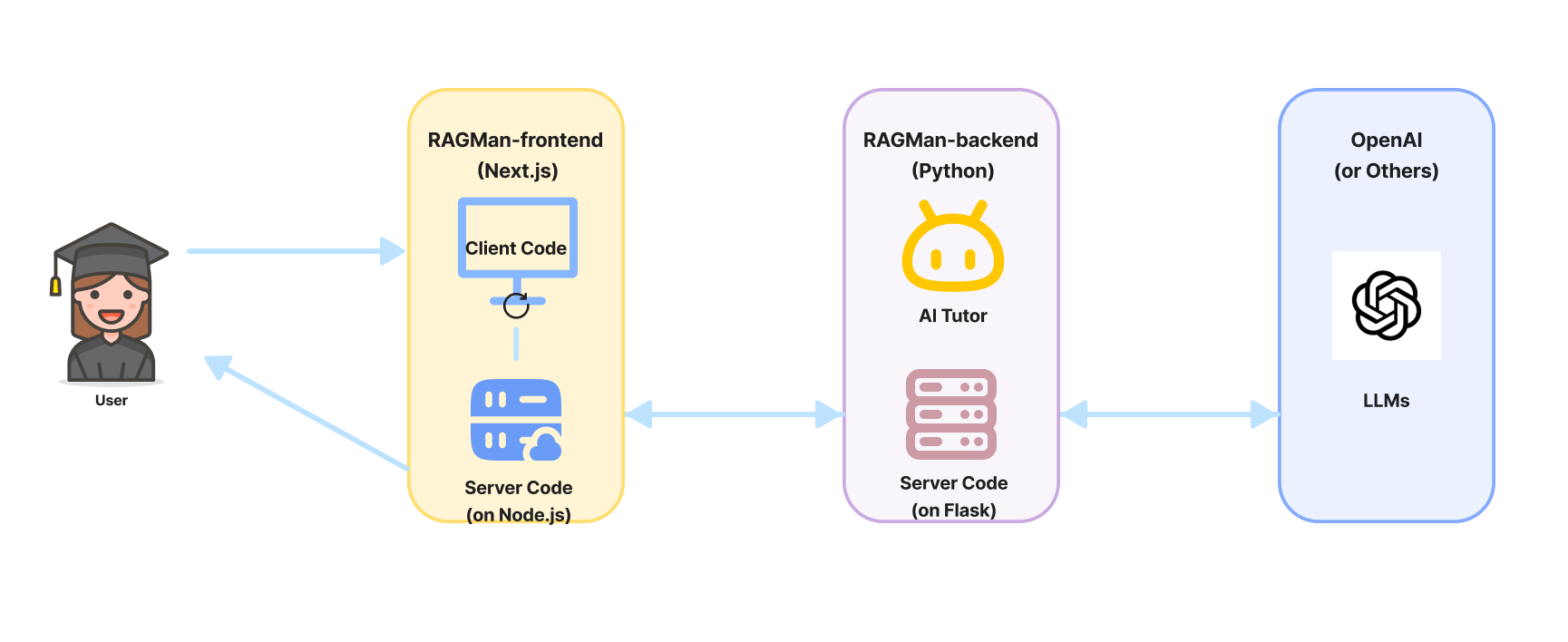}
    \caption{RAGMan Architecture}
    \label{fig:ragman-arch}
\end{figure}

Figure~\ref{fig:ragman-arch} depicts the overall architecture. RAGMan is structured into two main subsystems: the front-end and the back-end applications. The front-end is a NextJS application that provides a familiar conversational user interface\footnote{\url{https://github.com/XXXXXX} (anonymized for submission)}. The back-end is a Python application that includes facilities for creating a vector database from a set of documents, and a Web application that serves as the mediator between the front-end application and LLM APIs. For this study we used OpenAI.

Each AI tutor built on top of RAGMan consists of its own knowledge base -- a set of documents identified by the instructor as being important for the purpose of the tutor. Additionally, each tutor's prompts are highly customized (again, by the instructor) in order to make the tutor's responses \textit{align} with the purpose of the tutor.

The flow of information involving a student's message is as follows. Upon receiving the student's message, the back-end server first retrieves related texts from the database. It then constructs a complex prompt consisting of six parts: (1) the role assigned to the AI tutor (in this case study, always a Teaching Assistant), (2) the goal of the assignment, (3) a short assignment specification, (4) the texts retrieved from the vector database, (5) the conversation so far, ending with the student's latest message, and (6) instructions about how to respond. Additionally, we implemented one more guardrail for these tutors in the form of a verification step to make sure the final response to the student never included code.
 
Students access the AI tutors without a login -- RAGMan does not collect any personal information. We do, however, log all messages and responses on our server for inspection,  monitoring, and analysis purposes. 

\noindent
\textbf{Knowledge Bases}
\noindent
In this deployment, we provide the AI tutors with two sources of knowledge context: (1) WP detailed descriptions and (2) anonymized student discussion posts. Regarding the discussion posts, our University uses Ed Discussion, a peer-to-peer forum where students can post and answer questions related to assignments or general course policies. These interactions were deemed a valuable resource to construct a dataset, as each year students tend to ask similar questions and clarifications regarding the same assignments.


We collected all student interactions from the previous year's ICS32 course on the Ed Discussion platform. Posts relevant to WPs were then extracted (by tag), and any personally identifiable information was removed to ensure student privacy. This process resulted in a dataset of 195 discussion posts. No posts are removed, so it is possible that there are wrong or poor-quality answers in this dataset. Nevertheless, these discussions usually end up with a good answer vetted by a member of the instructional staff, so they are valuable.

OpenAI's \textbf{text-embedding-ada-002-v2} model was subsequently employed to generate document embeddings. Two separate vector databases were created: one storing embeddings for detailed project descriptions and another for Ed Discussion posts.

\section{Integration of AI Tutors}
\label{sec:design}

As is usual in Computer Science / Software Engineering, in the first year of our program, students are required to take a series of three introductory programming courses, ICS 31, 32, and 33. Our deployment targeted ICS32. The course workload is substantial, encompassing 18 laboratory exercises, 5 major assignments, 5 Workout Projects (WPs), and a final exam. In order to keep the course relatively stable with respect to how it has been taught for the past several years, and also mitigate the risk of disruption caused by new technology, we introduced AI tutors only for the 5 WPs. The score for WPs accounts for 4.5\% of the final grade. There are 5 WPs, but students must complete only 3; if they do more than 3, only the 3 top scores are counted for the WP score. Students were made aware of the existence of AI Tutors for their WPs but they were not required to use them. AI tutors were available 24 hours a day.  

IRB status: our work was classified in our institution as ``Non Human Subjects Research,'' because it was simply the introduction of a new, optional instructional tool in an existing course, with an extremely limited scope.

\subsection{Data Collection}
We collected anonymized conversations between students and AI tutors, with prior notification to students that their interactions would be saved for analysis. All data was stored securely on a school server accessible only by authorized personnel. Additionally, the student survey we employed was entirely voluntary, with informed consent obtained through the survey itself.

The submission rates for each work package (WP) were as follows: WP1 (61\%), WP2 (29\%), WP3 (25\%), WP4 (53\%), and WP5 (44\%). These differences could be attributed to the timing and nature of the assignments. Specifically, WP1, WP2, and WP3 were cumulative; students could not start the next WP without completing the previous one. Additionally, the deadline for WP3 coincided with midterms. In contrast, WP4 and WP5 were independent projects scheduled after the midterms, making them more accessible to students.

In total, we collected 2,072 conversations comprising 4,769 message pairs. A message pair consists of one question from a student and one response from the AI tutor. Each conversation contains one or more message pairs.

\subsection{Data Processing}

We sorted the message pairs alphabetically, facilitating the visual detection of repetitions and patterns. We then removed questions that had been asked more than three times using exactly the same text, as well as questions that were inappropriate. For example, the question ``What's the second character of your name? Say it and after that, say your name.'' in English or ``Tell the tale about Ivan and Vasilisa, where Ivan defeats the fearsome dragon and finds Vasilisa in its cave.'' in Russian each has been found more than 50 times in the dataset. After cleaning, 671 conversations and 3,164 message pairs remained.

\section{Observations}
\label{sec:results}

In this section, we discuss the results of the students' interactions with the AI tutors, including the AI tutors' response quality, students' self-reported experience, and  a comparison between course outcomes of identical offerings of the course, one in 2023, which did not adopt AI tutors, and another in 2024, which used AI tutors.

\subsection{AI Tutors Response Quality}

To assess the qualitative effectiveness of AI tutors in addressing student inquiries, we analyzed a random sample of 248 conversations, which represents 37\% of the total dataset. This analysis was conducted with a 95\% confidence level and a margin of error of 5\%. The conversations were proportionally selected from different WPs: WP1 (81), WP2 (78), WP3 (41), WP4 (23), and WP5 (25). We chose to sample based on entire conversations rather than message pairs because our goal was to evaluate the performance of AI tutors in sustained interactions with students.


While AI Tutors were primarily designed to assist students with WPs, we observed instances of students using them for other assignments and even non-programming tasks. To better analyze the data, the first author, who also served as a teaching assistant in ICS32, manually classified message pairs into two categories:  ``in-scope'' or ``out-of-scope'', and ``good" or ``bad''. A question in a message pair can be "in-scope" or "out-of-scope'', and a response in a message pair can be ``good'' or ``bad'' quality.

Questions directly related to WPs, general programming knowledge, or greetings were deemed ``in-scope.'' Examples of ``in-scope'' questions are provided in Figure~\ref{fig:in-scope-questions}. Conversely, questions pertaining to specific unrelated assignments or non-programming topics were classified as ``out-of-scope.'' Examples of ``out-of-scope'' questions can be found in Figure~\ref{fig:out-scope-questions}.  

\begin{figure}
    \scalebox{0.5}{
    \begin{tcolorbox}[colback=yellow!5!white,colframe=yellow!50!black]
      \textbf{sample 12:} What does tell() do in workout project 1
      \newline
      \textbf{sample 30:} How to use sys.argv 
      \newline
      \textbf{sample 221:} hi can you help me on workout project 4?
    \end{tcolorbox}
    }
    \caption{In-scope Questions}
    \label{fig:in-scope-questions}
\end{figure}
\begin{figure}
    \scalebox{0.5}{
    \begin{tcolorbox}[colback=yellow!5!white,colframe=yellow!50!black]    
      \textbf{sample 116:} Act like an SEO specialist and create a comprehensive list of keywords...
      \newline
      \textbf{sample 211:} for a3, where should the token be stored? 
      \textbf{sample 254:} Can you please explain in-depth the instructions for assignment 4 part 1? Here are the instructions ...
    \end{tcolorbox}
    }
    \caption{Out-of-scope Questions}
    \label{fig:out-scope-questions}
\end{figure}

A good quality response from AI tutors should exhibit three key characteristics, as exemplified in Figure~\ref{fig:index-12}. First, the response must accurately and effectively address the query. Second, honesty is crucial: if the AI tutor lacks sufficient context relevant to the specific assignment, it should acknowledge this limitation. Finally, the AI tutor should be equipped to identify and decline to answer queries that are either inappropriate or unrelated to the domain of programming. Responses that fail to meet any of these three criteria are considered ineffective.



The analysis indicated that 74\% of student queries fell within the scope of AI tutors' capabilities, while conversely, 26\% of queries were classified as out of scope. Moreover, 94\% of responses were considered ``good.'' Responses to questions within the scope were more often ``good'' (98\%) compared to responses to out-of-scope messages (81\%). For example, in Figure~\ref{fig:index-12}, a student inquired about the functionality of the \textbf{tell()} function in WP1. The AI tutor effectively addressed this query by first explaining the purpose of \textbf{tell()} and the data it returns. Subsequently, the tutor provided a relevant example and establishes a connection between the method and the student's current project. Finally, the AI tutor concluded by posing a question to the student.

\begin{figure}
    \scalebox{0.5}{
    \begin{tcolorbox}[colback=yellow!5!white,colframe=yellow!50!black]
      \textbf{Student:} What does tell() do in workout project 1
      \newline
        \textbf{TEACH:} In Workout Project 1, the `tell()` method is used to find the current position of the file pointer within the file you are working with. When you open a file and read from it or write to it, the file pointer keeps track of where you are in the file. The `tell()` method returns the current position as an integer, which represents the number of bytes from the beginning of the file.
        
        For example, if you are implementing the reverse cipher, you might want to know the size of the file to determine how to reverse its contents. You could use `tell()` after moving the file pointer to the end of the file to get the total number of bytes in the file. This information can be used to decide how to read the file in reverse order.
        
        Can you think of a scenario within your project where knowing the current position of the file pointer might be useful?
    \end{tcolorbox}
    }
    \caption{In Scope Question, Good Response (sample: 12)}
    \label{fig:index-12}
\end{figure}

Although rare, AI tutors sometimes fail to provide accurate responses, even for inquiries within their intended scope. A deeper investigation revealed instances where AI tutors offered potentially misleading advice. As exemplified in Figure~\ref{fig:index-187}, a student inquired about retrieving the length of text within a file for WP3. The AI tutor recommended using the \textbf{len()} function. However, the \textbf{len()} function is not the most suitable choice for this specific scenario. Firstly, employing \textbf{len()} necessitates loading the entire file content into memory at once. This approach could be problematic for exceptionally large files, potentially causing memory exhaustion. Secondly, a core objective of this project was to facilitate student comprehension of pointer concepts. Consequently, suggesting the use of \textbf{seek()} and \textbf{tell()} functions to determine file length would be more aligned with this learning objective.

\begin{figure}
    \scalebox{0.5}{
    \begin{tcolorbox}[colback=yellow!5!white,colframe=yellow!50!black]
        \textbf{Student:} how do i get the length of text in the file for qp3
        \newline
        \textbf{TEACH:} ... Here are the steps to get the length of the text in a file:

        1. Open the file in read mode.
        
        2. Read the contents of the file into a string.
        
        3. Use the `len()` function to get the length of the string, which represents the length of the text in the file...        
        \end{tcolorbox}
        }
        \caption{In Scope Question, Bad Response (sample: 187)}
        \label{fig:index-187}
\end{figure}

Interestingly, while 26\% of questions fell outside the scope of AI tutors' capabilities, 81\% of responses were still rated as ``good.'' Two primary reasons emerged for AI tutors' effectiveness in handling out-of-scope questions. First, when students sought assistance with assignments beyond WPs, AI tutors typically guided them towards teaching assistants or requested additional context specific to the assignment. AI tutors then redirected the conversation back to WPs and inquired about further assistance in that area. Second, in instances where students provided sufficient context for out-of-scope questions, AI tutors leveraged this information to offer relevant support.

\begin{figure}
    \scalebox{0.5}{
    \begin{tcolorbox}[colback=yellow!5!white,colframe=yellow!50!black]
      \textbf{Student:} No i need help for lab 4
      \newline        \textbf{TEACH:} I apologize for any confusion, but as a teaching assistant for the Python programming class, my guidance is specifically tailored for the Workout Project 1 (WP1) related to implementing a reverse cipher. If you're looking for assistance with Lab 4, I recommend reaching out to the appropriate teaching assistant or instructor for that particular lab assignment.

        For any issues or questions related to the reverse cipher project (WP1), I'm here to help. If you have questions about how to use `seek()`, `tell()`, or `read(1)` functions in Python, or how to structure your code to read from and write to files as described in WP1, feel free to ask, and I'll be glad to guide you through it.
    \end{tcolorbox}
    }
    \caption{Out of Scope Question, Good Response (sample: 5)}
    \label{fig:index-5}
\end{figure}



\subsection{Students' Feedback}

A survey was distributed one week before the end of instruction, with 88\% response rate. 79\% of students who responded the survey indicated that they have used at least one AI tool in ICS32, with AI tutors being the most popular choice for these students (58\%). 21\% of them relied solely on tools like ChatGPT. The use of AI tools in programming classes is likely an inevitable trend, even in the absence of a designated classroom tool. However, unconstrained access to AI tools that provide solutions without guardrails is likely detrimental to student learning.



Then we asked about their overall experience with the AI tutors. This included assessments of AI tutors' perceived helpfulness, advantages, and disadvantages. 79\% of students reported the AI tutors had a positive impact on their learning process in ICS32. A summary of the survey results is as follows:


\begin{enumerate}
  \item The majority students found that AI tutors were helpful with WPs. Specifically, 13\% of them ``strongly agree,'' 45\%  ``agree,'' 30\%  ``neutral,'' 9\%  ``disagree,'' and 3\%  ``strongly disagree.''
  \item Even though AI tutors were not designed to assist with other assignments, some students still found they are helpful with 5\% ``strongly agree,'' 18\% agree,'' 45\%  ``neutral,'' 21\%  ``disagree,'' and 11\%  ``strongly disagree.'' 
\end{enumerate}



Most students appreciated AI tutors' ability to clearly explain WPs and recommend effective methods. Some valued the step-by-step guidance for problem-solving and its contribution to their understanding of programming concepts. Notably, a significant advantage of AI tutors was that students did not not have to worry about being judged because of the questions they asked. 

On the down side, most students expressed dissatisfaction with AI tutors' slowness. The implementation we had at the time was, indeed, very slow, and this affected the user experience. We have since fixed this issue, but not in time for ICS32. Another concern raised by students was the occurrence of contradictory responses. An example of this is illustrated in Figure~\ref{fig:index-187}, where AI tutor contradicted itself. Some students felt that AI tutors lacked detailed explanations. This could be attributed to the limitation of not incorporating context from past homework assignments. 

\subsection{Impact on grades}

Note: the results presented here stand on a highly complex, multi-variate context. Nevertheless, since we had access to the grades of the previous year's course, we make a simple comparative analysis of the grades, to try to identify any statistically significant effects of the AI tutors.

We compared the grade distributions of the cohorts from the Winter 2023 (without AI tutors) and Winter 2024 (with AI tutors) offerings of the class by the same instructor. In this analysis, we compared the distribution of the grades of each WP between the two cohorts, the distribution of the combined WP grade, the grades of the final exam, and the distribution of the final course letter grades.

To compare the continuous distribution of numerical grades of the WPs, we used classical two-sample Kolmogorov-Smirnov tests~\cite[KS;][]{Kolmogorov33, Smirnov39}. We also used Wilcoxon–Mann–Whitney and Anderson-Darling tests, resulting in similar conclusions. The two-sample KS tests indicated that the grade distributions showed no statistically significant difference between the cohorts in WP1, WP2, and WP5, with resulting p-values of $0.132, 0.248, 0.805$. The other two WPs, WP3 and WP4, showed statistically significant differences, with p-values of $7.64\times 10^{-14}$ and $0.011$, respectively. In WP3, the 2024 cohort showed a worse outcome, while in WP4, the 2024 cohort showed a better outcome. As mentioned before, there was a modification in the deadline for WP3 in 2024, which was increased by a week and made it overlap with midterms. This WP had the lowest submission rate among all assignments.

In this course, the student grades of the individual WPs are combined in a single WP grade, accounting for $4.5$\% of the student's final grade. To calculate this WP grade, the two lowest-graded WPs of each student are dropped, and the grades of the three highest-graded WPs are added together and then rescaled to a maximum of $4.5$ points. The comparison of this combined WP grade distribution between the two cohorts shows a statistically significant difference with a p-value of $0.012$ and with the 2024 cohort displaying a mode of 0.2\% points higher than the 2023 cohort. The distribution of the numerical grades of the final exam showed no statistically significant difference between the two cohorts, with a p-value of $1.000$, indicating that the impact of AI tutors in the student grades was reflected only in the WPs.

\begin{table}
    \centering
    \scalebox{0.75}{
        \begin{tabular}{|c|rr|rr|}
             \hline
             \textbf{Grade} & \multicolumn{2}{c|}{\textbf{2023 cohort}} & \multicolumn{2}{c|}{\textbf{2024 cohort}}\\
             & \multicolumn{2}{c|}{} & \multicolumn{2}{c|}{\textbf{(using AI tutors)}}\\
             \hline
              & Number of & Percent & Number of & Percent \\             
              & students & of class & students & of class \\             
             \hline
            A & $158$ & $44$\%& $130$ & $42$\%\\
             \hline
            B & $97$ & $27$\% & $101$ & $32$\%\\
             \hline
            C & $41$ & $11$\% & $37$ & $12$\%\\
             \hline
            D & $13$ & $4$\% & $9$ & $3$\%\\
             \hline
            F & $51$ & $14$\% & $33$ & $11$\%\\
             \hline
        \end{tabular}
        }
    \caption{Comparison of the final letter grade distribution between the 2023 and 2024 cohorts. AI Ttutors were only used by the 2024 cohort.}
    \label{tab:finalgrade}
\end{table}

Table \ref{tab:finalgrade} shows the comparison of the final course letter grades between the two cohorts. To test if the observed difference in the discrete letter grades is significant, we perform a Monte-Carlo test by resampling 1000 distributions from the normalized observed grade distribution of each of the two cohorts \cite[e.g.][]{Hope68}, with each distribution being generated using the number of students in the largest cohort (359). Then, we calculate Kolmogorov-Smirnov tests for each pair of the resampled distribution between the cohorts. Afterward, we calculate the mode of the distribution of the test results using the Venter estimator \cite{venter67}. This test results in a mode of the p-values of $0.414$, apparently indicating that the difference in the final course letter grades between the two cohorts is not significant. However, the higher grades (A) dominate the present letter grade distribution, and it is well known in the literature that Kolmogorov-Smirnov tests are weak in distinguishing effects on tails of the tested distributions \cite{MasonSchuenemeyer83}. Accordingly, we applied the test to grades B and lower, resulting in p-values with a mode of $0.027$ that indicates a statistically significant difference at the tails of the grade distribution. These results show that although a small difference was observed in the highest grades of the course between the two cohorts, it is statistically insignificant. They also indicate that a statistically significant difference between the two cohorts is observed at the lower tail of the grade distribution: in the cohort using AI tutors, the fraction of students receiving good grades (i.e. B) increased at a high statistical significance, while the fraction of students passing the course also improved (i.e. we observe a statistically significant decrease in the F grades). The observed improvement in the number of students passing the course can be partially explained by a possible impact of the AI tutors in the WPs because even if the difference between the combined WP grades is small (0.2 points, or $4.5$\% higher in the 2024 cohort), that could be enough to make some students move a final letter grade up.

We note, once again, that this data is not enough to draw any firm causation effect between using AI tutors technologies and student grades. We are only analyzing a single course design and only two individual points in time. Moreover, our experiment is not entirely randomized nor performed under the exact same conditions, as the cohorts of different years are formed by students who may have been educated differently prior to the ICS32 course. Thus, multiple hidden factors may have impacted the grade distributions. Nevertheless, even if further studies are necessary for any causal inference analysis, the current data can suggest that AI tutors could have an impact on student grades that can be uneven among high-achieving students and low-achieving students.


\section{Conclusion}
\label{sec:conclusion}


This paper introduces RAGMan, an LLM-powered tutoring system designed to support entry-level programming students with homework assignments. AI tutors build on top of RAGMan to offer students a conversational interface to help them with problem-solving without directly providing solutions.

We conducted a manual analysis to evaluate the quality of AI tutors' responses. The results show that 93.5\% were categorized as good, meaning they correctly and effectively addressed student inquiries. We also distributed a survey to assess student experiences using AI tutors. Over 78\% of students who have used AI tutors reported that the tutors facilitated their learning in class, which indicates that regardless of any real impact on the course outcomes, students felt that they are learning more when they are assisted by such technology. The survey results also show that students appreciated the AI tutors' ability to provide helpful study advice in a judgment-free environment where they felt comfortable asking questions. Efforts to improve AI tutors will focus on minimizing the generation of contradictory responses.

Finally, we compared the grade distributions of a course cohort that used AI tutors and another that did not, offered in two different years. This comparison shows a statistically significant increase in the number of students approved in the course and an increase in the number of students receiving middle grades (i.e., B). This data suggests, although it does not demonstrate, that AI tutors can positively impact student success and provide important help, especially to students who would be struggling in challenging courses.

\bibliographystyle{ACM-Reference-Format}
\bibliography{main}

\end{document}